\documentclass{pasj00}

\begin{document}
\SetRunningHead{Deguchi et al.}{Kinematics of  Red Variables I. Basic Data}
\Received{2011/04/11}
\Accepted{2011/08/09 ; PASJ Ver.2.1 ;  Aug. 08, 2011}

\title{Kinematics of Red Variables in the Solar Neighborhood  I . Basic Data Obtained by an SiO Maser Survey}

\author{Shuji \textsc{Deguchi}\altaffilmark{1,2},  Tsuyoshi \textsc{Sakamoto}\altaffilmark{3}, }
\and
\author{Takashi \textsc{Hasegawa}\altaffilmark{4}}

\altaffiltext{1}{Nobeyama Radio Observatory, National Astronomical Observatory,\\
              Minamimaki, Minamisaku, Nagano 384-1305}    
\altaffiltext{2}{Graduate University for Advanced Studies, 
National Astronomical Observatory, \\
Minamimaki, Minamisaku, Nagano 384-1305}
\altaffiltext{3}{Bisei Space Guard Center, 1716-3 Ookura, Bisei, Ibara, Okayama 714-1411}
\altaffiltext{4}{Gunma Astronomical Observatory, 6860-86 Nakayama, Takayama, Agatsuma, Gunma 377-0702}



\maketitle

\begin{abstract}
In order to study the streaming motions of miras in the Solar neighborhood,
we newly surveyed  379 red variables in the SiO maser lines at 42.821 and 43.122 GHz 
with the Nobeyama 45m radio telescope.
Accurate radial velocities were obtained for 229 (220 new) detected stars.  
The sample  is selected from optical variables found by new automated surveys:
 the Northern Sky Variability Survey and the  All Sky Automated Survey.
The new sample consists of the "bluer" objects compared with those observed in the previous SiO surveys.
The distances to the objects are estimated using the period-luminosity relation, 
and they are mostly less than 3 kpc from the Sun.
The longitude-velocity diagram reveals three prominent  groups of  stars deviant from the circular Galactic rotation
with a flat rotation curve. In addition to the Hercules group of stars which was studied before,
we found two new deviant groups: one toward the Perseus arm and the other toward the Sagittarius arm. 
These two groups both exhibit anomalous motions toward the Galactic center, which seem to be
consistent with the noncircular motions of these spiral arms found in the recent VLBI proper-motion measurements 
for maser gas clumps. 	 
\end{abstract}

\section{Introduction} 
Moving groups are clumps of stars sharing the same spatial motion in the Solar neighborhood. 
They are often considered to be a fossil,  which keeps past dynamical information after its birth in the Galaxy. 
The coherent spatial motions of the moving groups are well studied in the past  based on
the Hipparcos  and the RAVE (the Radial Velocity Experiment; \cite{zwi08}) databases  (see \cite{fam05}).
In particular, the Hercules group of stars, which was first identified by O. J. Eggen (see a summary by \cite{egg96}), is a well studied 
moving group with rotational lag and outward motion of about 40  km s$^{-1}$ and 50  km s$^{-1}$, respectively,  
to the Galactic rotation. It is inferred that  a few percent of stars in the Solar neighborhood are members of this group \citep{ben07}.
The origin of the Hercules group is attributed to a rotational resonance of the bar-like Bulge, because
 the population of stars of this group  is a mixture with different ages \citep{ben07}.  
\citet{fea00} investigated an outward motion of short-period Mira variables near the Sun,
and attributed it to the  resonance effect of the Bulge bar.  Presumably
the Hercules moving group, which was found in the Solar neighborhood,   spreads spatially far from the Sun.
\citet{deg10} found that a group of maser stars
in the Galactic longitude range between 20$^{\circ}$ and 40$^{\circ}$,
 which are located at a few kpc from the Sun,
 have a distinctively large outward motion 
compared with the motions  of usual stars under the Galactic rotation. They also
attributed the large outward motion to the effect at the outer Lindblad and corotation resonances of the central bar. 
The resonance effect of the Galactic bar should appear in areas
near the resonance circles in the Galactic plane. In  particular, the old stars with ages several times longer than the 
rotational period of the bar pattern reflect the resonance effect. 
 Therefore, miras are ideal sample for studying the bulge-bar resonance effect because they are evolved stars with ages of about a few Gyr.
In contrast,  tidal streams of dwarf galaxies, e.g., the Sgr dwarf stream (for example, \cite{maj03}),  are often traced
in a relatively limited area of the sky far from the Galactic plane because of their low stellar density, 
though they still have been found  in the Galactic disk using  blue metal-poor stars (e.g., \cite{bel07}). 

Radial velocities of  OH and SiO maser sources have been used to investigate dynamics 
of stars in the disk \citep{jia96,nak03,ita01} and the bulge of the Galaxy \citep{izu95,sev01}.
These maser stars are mostly miras and semi-regular variables, i.e.,  O-rich evolved stars
at the asymptotic-giant-branch (AGB) phase, though a small amount of red supergiants are contaminated in the sample.
Previous surveys of these stars by OH and SiO maser lines were preferentially made 
for the highly reddened, optically very faint stars because of their high detection rate in a color-selected sample 
(for example, see \cite{deg04}).
These stars are located at relatively large distances in the Galactic disk, 
compared with a sample of optical miras. Therefore,
optical miras in the Solar neighborhood are missing 
in the previous samples of maser sources [e.g., \citet{nak03,deg07}, except \citet{jew91}]. 
Even though radial velocities have been obtained for a large number of optical miras  
 by optical spectroscopy, the accuracy in the measurement is quite limited. 
 For example, if we compare the radial velocity of a mira in the RAVE database
 with that of OH or SiO masers, we often find typically
 a 10 km s$^{-1}$ or much larger  difference between them. 
 This is caused by several reasons:
 insufficient spectral resolution in optical instruments,
phase dependency of optical line velocities on mira pulsation (see, e.g.,  \cite{sch00}),  
and a velocity shift of the optical lines due to scattering by moving circumstellar dust
\citep{van82}. It is known that the stellar velocities obtained in the maser line measurements are 
accurate within $\sim 2$ km s$^{-1}$ [e.g., see section 3.1 of \cite{nak06}].
Therefore, it is useful to measure the radial velocities of optical miras 
in the Solar neighborhood in SiO maser lines 
even if optical velocities are available for some stars. 
Moreover,   the accurate radial velocities  of miras in the Solar neighborhood by the maser observations are essential
to see if the deviant stream is continuously connected with the stream that is found previously in a large extension of the Galaxy
 \citep{fea00,deg10}. 
In addition, since precise measurements of proper motions will be available
by the phase-reference VLBI technique for SiO maser sources \citep{kob08},
 accurate 3d motions in space will reveal  in future for these objects. 
 
In this paper, we present the result of  a new survey of the optical red variables  
in the SiO maser lines with the 45m telescope at Nobeyama.
A number of new variable stars were recently found by automated optical variability surveys: 
the Northern Sky Variability Survey (NSVS;  \cite{woz04}) and the All Sky Automated Survey (ASAS; \cite{poj05}).
Though these newly found optical variables are much bluer in near-infrared (NIR) colors (such as $H-K$) than the typical SiO maser sources  previously surveyed,
they exhibit the characteristic optical variability of miras.
Since the bluer color indicates the higher surface temperature and smaller mass loss rate 
of the central star  in general,  the detection rate of SiO masers was expect to be very low  for such a sample. 
However,  contrary to our expectation, our preliminary survey made in 2009 resulted in 
a significantly high  detection rate of SiO masers. 
Therefore, we have performed a new extensive observation in the SiO maser lines toward these red variables, 
and have increased the data of our SiO radial velocity database. 
In this paper, we present the result of the observations and give a limited discussion on the kinematic properties of this sample, 
based mainly on the radial velocities.
For all of the sampled stars, proper-motions have been measured optically \citep{roe10}. 
A kinematic study based on the proper motions will be given in the future paper. 
 
\section{Observation, sample selection, and results}
\subsection{Observation}
The observations were made with the 45m radio telescope at Nobeyama in 2009 March, 2010 March--May, and 2010 December--2011 January 
in the SiO $J=1$--0 $v=1$ and 2 transitions at 43.122 and 42.821 GHz, respectively.
A few data  taken before 2009 were also added for the present analysis.
A cooled HEMT receiver (H40)  was used for the 43 GHz observations
with acousto-opt spectrometer arrays with 40 and 250 MHz bandwidths (with velocity resolutions
of about 0.3 and 1.8 km s$^{-1}$, respectively).
The system temperature was about 180 --- 250 K for the SiO observations,  depending on weather conditions.
The half-power beam width (HPBW) of the telescope was about 40$''$ at 43 GHz. 
A conversion factor of the antenna temperature to the flux density was about 2.9 Jy K$^{-1}$. 
All of the observations were made by the position-switching mode. 
Further details of observations using the NRO 45-m telescope have been
described elsewhere (see \cite{deg00}). 
The spectrometer arrays also covered the SiO $J=1$--0 $v=0$ and $v=3$  lines
at 43.424 GHz and 42.519 GHz, respectively, the $^{29}$SiO  $J=1$--0 $v=0$ line at 42.880 GHz,
and H53$\alpha$ at 42.952 GHz. However, these lines were detected in a few  sources
(shown in Appendix 1). 
\begin{figure*}
  \begin{center}
    \FigureFile(110mm,80mm){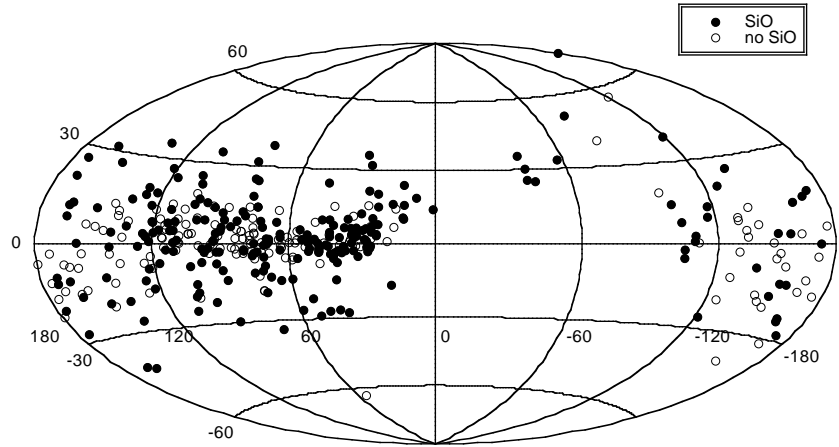}
  \end{center}
\caption{ Distribution of the observed objects in the Galactic coordinates in the Hammer-Aitoff projection. 
Filled and unfilled circles indicate SiO detection and no detection, respectively. 
}\label{fig: aitoff}
\end{figure*}

\subsection{Sample selection}
The sample for the present SiO maser searches was chosen 
mainly from the "Catalog of Red Variables in the Northern Sky Variability Survey"\citep{wil04}.
Because the coverage of this survey is heavily weighted on the northern Galactic plane,
we used an additional optical catalog of red variables selected from the "ASAS Variable Stars in Southern Hemisphere"
 \citep{poj05}.  These two catalogs  listed up the red variables found in automated sky surveys. They give  period of
 optical light curve, classification code, optical magnitude and amplitude, coordinates of the stars with accuracy better than 10$''$, and
 2MASS and IRAS identifications. 
From these catalogs, we selected the objects 
 with a classification code of "M" (mira) or "SR+L"
 (semiregular \footnote{The variability type,  "semi-regular", is applied to the variables with smaller amplitude, shorter periods, 
 and more irregular pulsations than miras; some occasionally show multiple periodicity \citep{bed98}.} 
 and irregular variables) and  with a period longer than 80 d 
 [which covers enough for SiO maser stars at the short-period limit ($\sim 150$ d)].
Additionally, we applied  the selection criteria to effectively squeeze out the stars 
 enshrouded by circumstellar dust;  $K<9$, and $H-K>0.6$,           
 the 12 $\mu$m flux density brighter than 3 Jy, and the color $-0.5 <C_{12}$ \ $  [\equiv log(F_{25}/F_{12})] \lesssim 0.2$, 
where  $H$ and $K$ are 2MASS $H$ and $K_s$ magnitudes, respectively \citep{cut03}, and  
$F_{12}$ and $F_{25}$ are the IRAS flux densities in the 12 and 25 $\mu$m bands, respectively  \citep{bei89} 
[the MSX bands C and E \citep{ega03} were also consulted for the $|b| \lesssim 6^{\circ}$ sources.
We applied the same criterion in $C_{12}$ by translating $log(F_E/F_C)$ to $C_{12}$ without any correction, 
where $F_C$ and $F_E$ are MSX 12 and 21 $\mu$m flux densities.
Detailed comparison of the MSX colors with those of IRAS \citep{sjo09} showed that the correction 
is negligibly small around $C_{12}=-0.4$ (see their Figure 4)].  
All the  objects in the present sample are optically identified variable stars. 
These are supposedly late-type (AGB or post-AGB) stars surrounded by a dust envelope  with a color-temperature range between 250 and 1200 K
 (they are the stars with a thin dust envelope in region II and IIIa of the mid-IR two-color diagram of  \cite{van88}). 
The stars cataloged as a carbon star in the SIMBAD database are excluded from the sample.
Furthermore, we selected a few bright objects for backup observations in a bad weather condition. They are slightly 
bluer sources in $H-K$ and in $C_{12}$ but have not been surveyed before. We added these additional objects to our results 
for completeness. We have observed all the red variables  in the \citet{wil04}'s catalog down to $F_{12}=7$ Jy
(though we could not consume all the bright objects in the ASAS catalog).
  The distribution of the observed stars in the sky is shown in Figure 1.
  
  \subsection{Results}
\begin{figure}
  \begin{center}
    \FigureFile(80mm,80mm){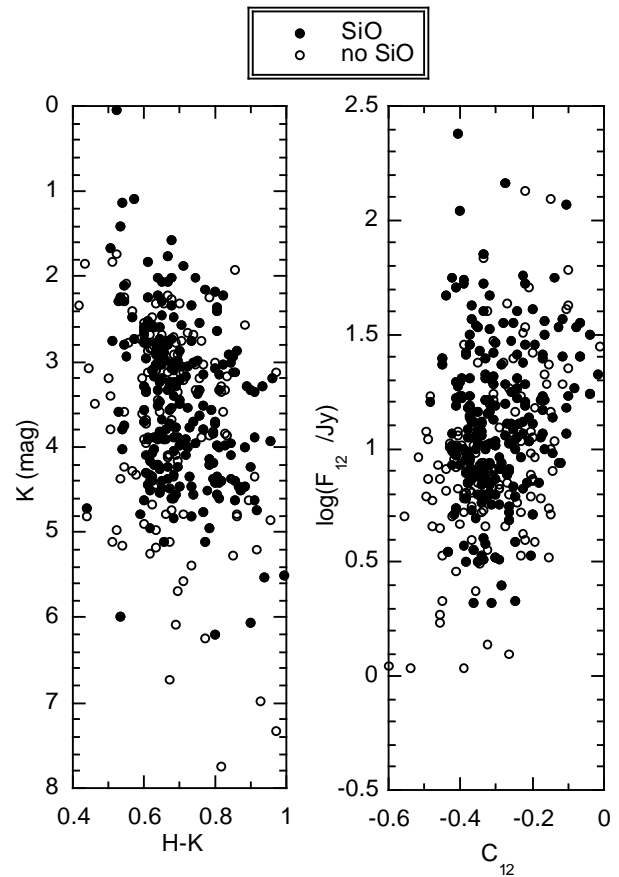}
  \end{center}
\caption{NIR  color-magnitude and and MIR color-flux density diagrams for the sampled objects.
Filled and unfilled circles indicate SiO detection and no detection, respectively. 
}\label{fig: magnitude-color}
\end{figure}

Observational results are summarized in Tables 1 and 2 for the SiO detections and no detections, respectively.
The observed spectra of the SiO $J=1$--0 $v=1$ and 2 transitions for detected sources are given in Appendix 1,
and the individually interesting objects are also discussed there.
Table 3 summarizes a few detections in the additional lines of SiO, i.e., the $^{28}$SiO $v=3$ and $v=0$ $J=1$--0 and
$^{29}$SiO $v=0$ $J=1$--0 transitions [the spectra are shown in Appendix 1].
Table 4 summarizes the infrared properties of all the observed sources.

Figure 2 shows the near- and middle-infrared color--magnitude 
diagrams for observed stars. 
If we compare this, for example, with Figure 2 of \citet{deg10},  we can recognize that the present sample is weighted 
toward bluer colors than the previous SiO maser survey samples; for example,
the median of $H-K$ for the present sample is  0.72, while it is 1.09 for the former sample,   
 and the median of  $C_{12}$ in the present sample is $-0.29$, while it is  $-0.16$ for the former sample.
In addition, the objects in the present sample are
much brighter in $K$ band than those in  the previous samples.  It  suggests that  
the average distance from the Sun of the sampled stars is much smaller than that of the previous SiO-survey samples
with a typical distance of $\sim 4$ kpc.
 The detection rate of SiO masers is quite high ($\sim 80$ \%) for the objects that are bright in the 12  $\mu$m and $K$ bands,
 but gradually  decreases as the infrared flux density decreases. Beyond $K=5$, the number of no detections exceeds
 that of detections because of the large distance to the sources. 
Such a high SiO detection rate in SiO maser emission was an unexpected result, which apparently does not match up with the blue colors of the sample.
However, this apparent discrepancy could be explained for the following reasons. In  the previous SiO surveys, the variability indices 
of the IRAS catalog were not considered in the selection criteria (except \cite{jia96}).
Therefore, the samples could include young steller objects (YSOs) and red giants (RGB stars) 
 which may mimic IR colors of the AGB stars , but do not emit SiO masers. On the contrary, the present sample is selected from the
 optically visible variable stars. It assures that they are stars in the AGB or post-AGB phases exhibiting active mass loss.
\begin{figure}
  \begin{center}
    \FigureFile(80mm,50mm){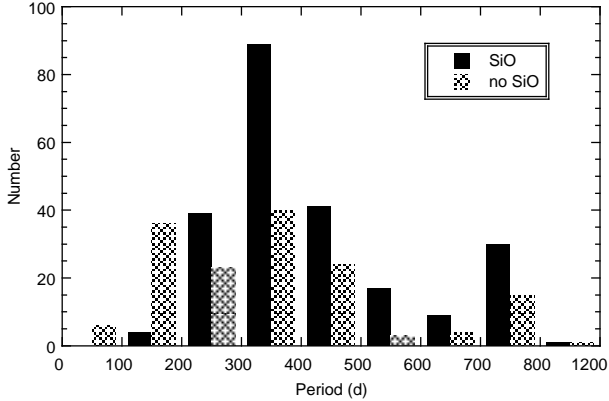}
  \end{center}
  \caption{Histogram of period for SiO  detections and no detections.
  The filled and shaded areas indicate the SiO detection and no detection, respectively.
}\label{fig: N-P}
\end{figure}
%
\begin{figure}
  \begin{center}
    \FigureFile(80mm,50mm){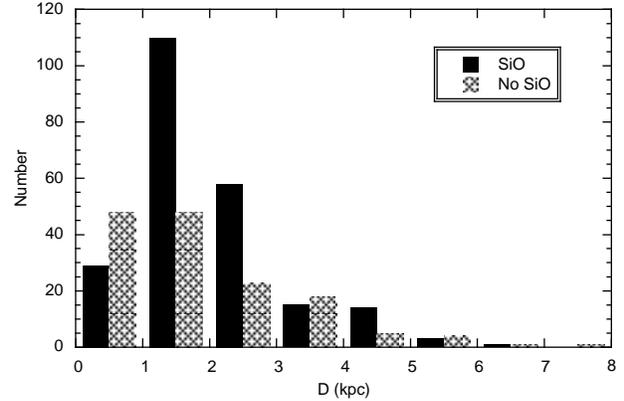}
  \end{center}
  \caption{Histogram of luminosity distances computed using the PL relation.
  The filled and shaded area indicates the SiO detection and non detection.
  The average distance is  $2.0\ (\pm 1.1)$ kpc for SiO detections and $2.2\ (\pm 2.5)$ kpc for no detections,
  where the parenthetic number is a standard deviation.   
}\label{fig: hist-distance}
\end{figure}
 
Figure 3 shows a histogram of period for the detections and no detections. 
The averaged period is 424 ($\pm 158$) d for the detections and 347 ($\pm 192$) d for the no detections,
where the parenthetic number is a standard deviation.  
The SiO detection rate seems to increase with period,
as has been found in the past surveys (e.g., see Figure 3 of \cite{deg04}).  
We found 4 SiO maser sources with a period shorter than 200 d :
J18424774+1548565, J19091839+7333285 (the shortest P=154 d),  J20414535+3353226, and J22325976+6654394.
All of them are semi-regulars.
The blue miras with a period less than 200 d occasionally exhibit a large deviant motion from the Galactic rotation \citep{fea00}.   
However, above 4 objects (distributing in the Galactic longitude and  latitude ranges of $l=46$ -- $110^{\circ}$ and
 $b=-5$ -- $+24^{\circ}$)  spread only in the velocity range between $-16$ and $25$ km s$^{-1}$. 
 Therefore, we do not find any anomalous kinematics for these 4 objects only from their radial velocities.

 We estimated distances to the observed stars  based on the PL (Period-Luminosity) relation \citep{whi08}.
 The detail of the distance estimation is given in Appendix 2. Figure 4 shows histogram of distances for the detections and no detections.
This figure indicates that most of objects in the sample are located within a distance of 3 kpc from the Sun except a few faint ones,
  though there is a considerable uncertainty in the distance estimation. The average distance is 
  2.0 ($\pm 1.1$) kpc for the detections, and 2.2 ($\pm 2.5$) kpc for the no detections.
  
\section{Discussion}
\subsection{Longitude-velocity diagram}
In Figure 5, we present the longitude-velocity diagram of the detections, in which filled and unfilled 
circles indicate  the galactic latitude ranges of  $|b|<10^{\circ}$ and  $|b|>10^{\circ}$, respectively.
Thick curves indicate the expected radial velocities for the objects under a circular rotation at distances, 1, 2, and 4 kpc from the Sun.
Here we assumed a flat Galactic rotation curve  of 220 km s$^{-1}$ in the Solar neighborhood 
and the Sun--Galactic-center distance of 8 kpc. 
We have also drawn the broken curves which are
expected for the Hercules and Arcturus moving groups of stars near the Solar neighborhood
(at the distance of 1 kpc).
For simplicity, we assumed  that the rotational lag and radial motion of the streams to the Galactic rotation
are kept  the same everywhere near the Solar neighborhood.   
The curve for each moving group strongly depends on the  assumed velocity law:
 see,  for example, Figure 8 and Appendix 3 of \cite{deg10}).

 In Figure 5,  we see notable concentrations of the stars with 
$|b|<10^{\circ}$ near the curve of 4 kpc distance;
one around $l=25$ -- 45$^{\circ}$ and  $V_{\rm lsr}\simeq +40$ -- +80 km s$^{-1}$,
and the other around $l=95$ -- 135$^{\circ}$ and  $V_{\rm lsr}\simeq -70$ -- $-40$ km s$^{-1}$.
In addition,  there is a  group of stars in the area $l=20$ -- 50$^{\circ}$ and $V_{\rm lsr}\simeq -80$ -- $-20$ km s$^{-1}$,
which is deviant from the Galactic rotation by more than $\sim 50$ km s$^{-1}$.
These deviant groups of stars are surrounded by ellipses in Figure 6 for clarity. 
They are overlaid on the CO $l$-$v$ map \citep{dam01} for comparison.

The negative velocity  feature around  $l=20$ -- 50$^{\circ}$  ($V_{\rm lsr}\simeq -80$ -- $-20$ km s$^{-1}$)
 has been discussed extensively by \citet{deg10}.
 This is likely an extension of the Hercules moving group of stars,
which is caused by outer Lindblad resonance of the Galactic bar structure.
\citet{bov10} predicted that the member stars of the Hercules moving group would reveal
 most promisingly in the  Galactic longitude range of 
$250^{\circ} \lesssim l\lesssim 290^{\circ}$.  Unfortunately, stars in this longitude range 
are difficult to observe from Nobeyama except for stars at high Galactic latitudes. 
Furthermore,  SiO maser sources (O-rich evolved stars) 
are not populated much outside the Solar circle (e.g., \cite{jia96}). Therefore, 
it is hard to confirm such a prediction 
through only the present discussion based on the longitude-velocity diagram.
\begin{figure*}
  \begin{center}
    \FigureFile(100mm,70mm){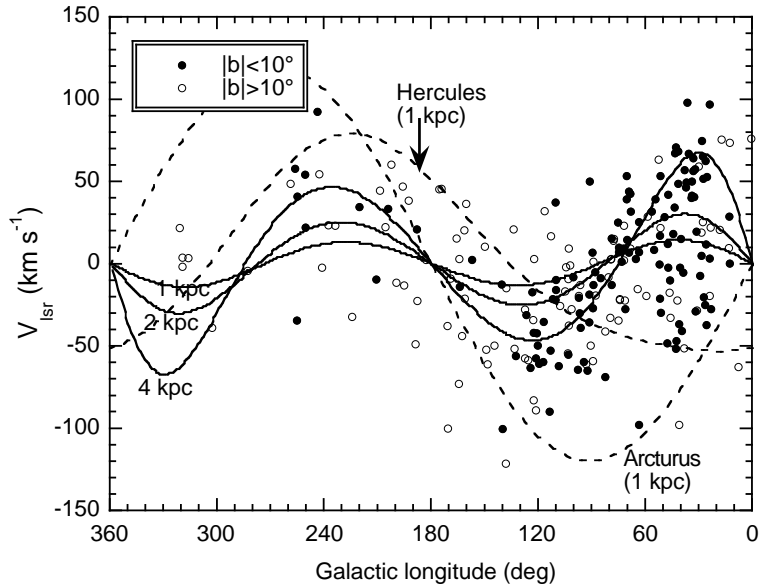}
  \end{center}
  \caption{Longitude-velocity diagram for SiO detected sources in the present sample.  
Filled and unfilled circles indicate objects below and above $|b|=10^{\circ}$.
 Three thick curves indicate radial velocities expected from the model with a flat rotation curve of 220 km s$^{-1}$  
 for stars at distances, 1 , 2 , and 4 kpc from the Sun, respectively. 
Broken curves indicate  radial velocities expected for the  Hercules and Arcturus moving groups 
 of stars  (with a rotational lag of $-42$ km s$^{-1}$ and an outward velocity of 52 km s$^{-1}$ for the Hercules group,
and a rotational lag of $-120$ km s$^{-1}$ for the Arcturus group) at a 1 kpc distance
 from the Sun. 
}\label{fig: l-v}
\end{figure*}
\begin{figure*}
  \begin{center}
    \FigureFile(110mm,70mm){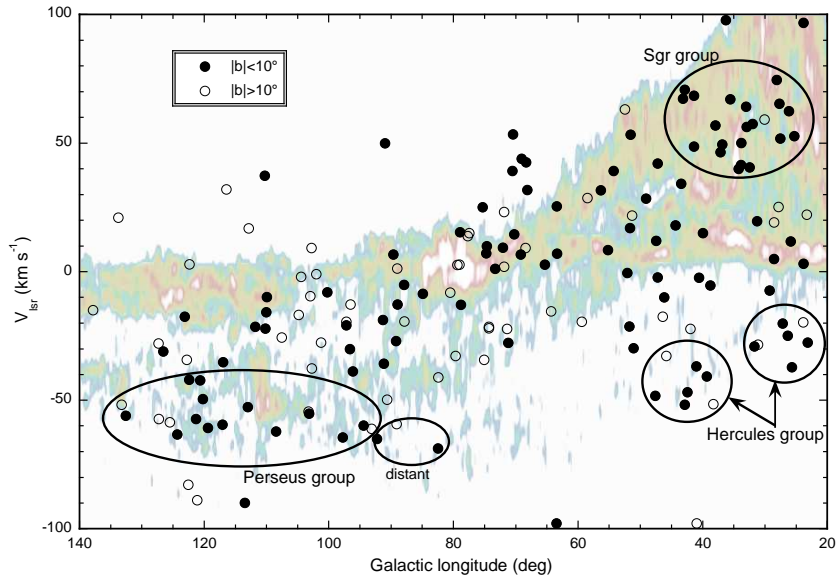}
  \end{center}
  \caption{A part of figure 5 but overlaid on the CO longitude-velocity map taken from \citet{dam01}.
  Filled and unfilled circles indicate the SiO sources below and above $|b|=10^{\circ}$.
 The large ellipses indicate the deviant groups of stars toward the Sagittarius and Perseus arms, and for the Hercules group of stars.
 Small ellipse assigned as "distant" indicates the distant stars which are not associated with  the Perseus deviant group.
}\label{fig: Sgr-Per-arm}
\end{figure*}

\subsection{Perseus group of deviant stars}
One of notable characteristics of Figure 5 is that a considerable number of objects exhibit radial velocities
larger than that expected from the 2 kpc distance; many stars fall around or beyond  the curve of the 4 kpc distance.    
Because the objects in the present sample are optical variables which are bright in $K$ band ($K<4$),
most of these are to be close to the Sun, i.e., their distances are smaller than 3 kpc. 
This discrepancy between the luminosity and kinematic distances is discussed below.

Figure 6  compares the distribution of the SiO sources with that of CO emission.
The concentration of the points seen in the area of  $l=95$ -- 135$^{\circ}$ and  $V_{\rm lsr}\simeq -70$ -- $-40$ km s$^{-1}$
coincides with a peak of  CO emission feature for the Perseus spiral arm (see \cite{dam01}).
CO emission is most prominent at around $l=111^{\circ}$ and $V_{\rm lsr}= -45$ km s$^{-1}$
 (toward  the NGC7538 molecular cloud). 
The average distance for the 18 objects in the ellipse of figure 5 (marked as Perseus) is estimated to be 
1.94 kpc (with a standard deviation of $\pm 1.10$ kpc) based on the period-luminosity (PL) relation. 
There is a large difference between the kinematic and luminosity distances for this group of stars. 
If we believe the luminosity distance,  the stars of this group are approaching us with velocity
larger than the velocity expected by the standard circular rotation of the Galaxy.
 Note that the SiO maser sources have a velocity dispersion of about 25 km s$^{-1}$ from the 
 average Galactic rotation [see the discussion in the last paragraph of Appendix 2 of \citet{deg05}].
 
 The distance to the Perseus spiral arm was controversial in the past \citep{rik68,rob72}.  
Recently parallax distances to the several masing objects in this spiral arm 
have been measured with Very Long Baseline Interferometric (VLBI) technique. 
For example,  the distance to W3(OH)  ($l=134^{\circ}$) is determined to be
$1.95\pm 0.04$ kpc \citep{xu06}.  A comprehensive summary of the objects with annual parallax measurements 
is found in Figure 11 of \citet{asa10}, which visualizes positions and peculiar motions of several objects in this spiral arm. 
The Perseus spiral arm exhibits a systematic deviation from the circular rotation
by a $\sim$30 km s$^{-1}$ in the longitude range $l=90$ -- 150$^{\circ}$.
Since the 18 red variables toward the Perseus arm exhibit a similar kinematic characteristic,   
we conclude that these variables are associated with the Perseus spiral arm.  
An average period of the 18 red variables in the Perseus arm is  423 ($\pm 109$) d,  which is 
longer than the average period  of field optical  miras of about 300 d \citep{whi00,tem05}. 
This fact indicates that they are relatively massive stars compared with the field miras.
For example, the initial mass of AGB stars with P=420 d is about 2.5 $M_{\odot}$ with age of 0.9 Gyr (see, e.g., Figure 20 of \cite{vas93}).
Thus, the red variables in the Perseus arm has not left far from the birth place.

 \subsection{Sgr group of deviant stars}
As well as the Perseus spiral arm,
we may consider a possibility of association of another deviant group at 
$l=25$ -- 45$^{\circ}$ and  $V_{\rm lsr}\simeq +40$ -- +80 km s$^{-1}$  with the Sagittarius-Crux arm.
The average distance and period of 21 stars in this group (the upper ellipse noted as Sgr group in Figure 6) are 2.58 ($\pm 0.99$) kpc and
489 ($\pm 160$) d, respectively.  Notable structures  in this direction  in our Galaxy  are the Local Spur ($D\sim 0$--2 kpc),
the Sagittarius-Carina spiral arm ($D\sim 2$ -- 3 kpc), and the Scutum-Crux arm ($D\sim 3$--5 kpc). 
It is possible that the Local Spur is a down stream branched from the Sagittarius-Crux arm. 
The branching to the Local spur seems to start
near the tangent of the Sagittarius-Crux arm at $l\sim 55^{\circ}$.

In the $l$-$v$ diagram (see the overlaid color map of figure 6),  
$^{12}$CO emission is very weak around 
($l$,  $V_{\rm lsr}$) $\sim$ (30$^{\circ}$, +60 km s$^{-1}$),
except in the direction of the HII region $G34.257+0.155$.
This is also true for the $^{13}$CO map \citep{lee01}. 
The kinematic distance of the HII region $G34.257+0.155$ was estimated to be 
3.8$^{+0.4}_{-0.8}$ kpc \citep{fis03} from the HI absorption feature 
assuming the standard circular rotation.
Recently the annual parallax distance was measured with VERA for the nearby infrared dark cloud, $G034.43+00.24$
 \citep{kur11}. The distance to the H$_2$O maser sources in this dark cloud is $\sim 1.56\pm 0.12$ kpc,
 which is considerably smaller than the kinematic distance of this cloud. 
 Because the radial velocity of this cloud ($V_{\rm lsr}\sim 50$ -- 60 km s$^{-1}$)
 is similar to that of the HII region $G34.257+0.155$, the distances of $G34.257+0.155$ may be overestimated.
 It is also likely that our deviant group of stars, which has a similar radial velocity
 in the same direction,  is likely to be in the same spiral arm.
Therefore, we call this group as the "Sgr" deviant group  because the average luminosity distance of the stars 
in this deviant  group is close to the estimated distance of the Sagittarius arm at $l\sim 30^{\circ}$ .
 
\citet{rei09} summarized the recent parallax measurements of  massive star forming regions
with VLBA and VERA by maser lines. Three star forming regions, G23.6$-$0.1, G35.2$-$0.7, and G35.2$-$1.7, 
exhibit the parallax distances in the range 
between 2 and 4 kpc in their table 4.  However, one of these ($G23.6-0.1$) has  a large radial velocity of $V_{\rm lsr}= +83$ km s$^{-1}$
(the kinematic distance $D^{Std}_{k}=5.04$ kpc or  $D^{Rev}_{k}=4.77 (\pm 0.3)$ kpc for their new rotational parameters),
 but the parallax distance of this object is 3.19 kpc, which locates this objects very near the far arm in this direction (the
Scutum-Crux arm; see figure 5 of \cite{rei09}).  Another two source, G35.2$-$0.7 and G35.2$-$1.7,
 have much smaller radial velocities ( $V_{\rm lsr}= +28$ and +42 km s$^{-1}$),  for which the parallax  distance roughly 
 agrees with the kinematic distance (2 -- 3 kpc). From these facts and the 1.6 kpc distance of the dark cloud $G034.43+00.24$
 \citep{kur11}, we conclude that the spiral arm  in this direction has a complicated velocity structure and a large noncircular motion.
 
The exceeding velocity of the Sgr group of stars to the galactic rotation suggests either (1) 
that these stars move faster than the rotational velocity given by the flat rotation curve,
or (2) that these stars move toward the galactic center (this inward motion causes the radial velocity increase in this direction).  
Because the directions of these two motions appear in opposite sense in the proper motion,  
VLBI observations of proper motions  of objects in Sagittarius arm can be a good test of above cases.
 
 Because the average period of this star group ($\sim 480$ d)  is considerably large compared with
 the average period of optical miras, they are relatively young objects
 compared with the field miras. Therefore it is likely that these stars are born in the Sagittarius arm and do not completely  
 depart from this arm yet.

\subsection{Correction in the distance  for long-period miras}
 Optical and infrared properties of the candidates for the deviant groups are summarized in Table 5;
it gives the 2MASS name,  period, variability type, $R$ magnitude, $R$ amplitude of the variability, 2MASS $K_s$ magnitude,
error in the $K$ magnitude, quality flag for the $K$ magnitude,  luminosity and kinematic distances.
It involves  a few stars with low photometric quality in the $K$ band in the 2MASS catalog. 
These objects have relatively large distances compared with the average value of each group. Therefore
the small average luminosity distances for these two groups of stars are not due to the objects
with poor photometric accuracy.  

In the previous sections, we used the  PL relation which was derived from the photometric measurements of the miras
with periods between 100 and 400 d \citep{whi08}, and extended the linear relation to the longer period up to 1000 d.
However,  it has been known that some miras with $P>400$ d, especially for OH/IR sources, 
lie above the linear extrapolation of the PL relation (e.g.,  see \cite{fea09}).  
This may cause an error in the distance estimation for the deviant group of stars.
Therefore, we also applied the different  PL relation for  the stars with $P>400$ d,
which was derived from the longer period miras in the Large Magellanic Cloud (LMC) \citep{ita11}. 
The detail of this correction was described in Appendix 3. 
The corrected distances (applied for all of $P>400$ d stars) are given in parenthetic number in the 8th column of Table 5.
In this case, the average distance is 2.2 ($\pm 1.3$) kpc  for the Perseus group of stars,
 and it is 3.5 ($\pm 2.1$) kpc for the Sgr group of stars, where the parenthetic number is the standard deviation. 
Therefore,  such correction does not influence much for the distance estimation for the  Perseus group,
but it increases the average distance considerably for the Sgr group, because the latter group involves 
many stars with $P>400$ d.  It is likely that the Sgr group of stars involve several very long period ($P=730$ d)
stars, which contaminate this sample. If we remove the 4 stars with $P=730$ d,
which is the upper boundary as a result of insufficient data in the NSVS survey \citep{wil04},
the average of the corrected luminosity distances of the Sgr group of 17 deviant stars is
2.6 ($\pm 1.2$)  kpc. 
The average of the  kinematic distances of this reduced set  is  3.5 ($\pm 0.7$) kpc.
The Student t-test gives a probability of 2 \% for these two averages
being produced by the same distribution function. In other words,  with 98\% confidence level, we can state  
that the average distances of these two sets are significantly different.  
In table 6, we summarized the average distance and standard deviation, 
and the t probability of the Student's t-test for the Perseus and Sgr (with smaller number) groups.
In summary,  the discrepancy between kinematic and luminosity distances for the Perseus and Sgr groups of stars
are not removed by the correction in the distance for the $P>400$ d,
though evidence is slightly weak for the case of the Sgr group.\footnote{ 
Of course, the smaller sets using only the miras with $P<400$ d give luminosity distances, 1.9 ($\pm 1.2$) kpc
and 1.8 ($\pm 0.6$) kpc for the Perseus and Sgr groups, respectively. 
The difference between luminosity and kinematic distances is statistically significant in the Student's t-test  for both groups, 
which are consistent with the previous result including the  $P>400$ d stars.
}

\setcounter{table}{5}
\setlength{\tabcolsep}{2pt}\footnotesize
\begin{table}
 \begin{center}
  \caption{Distance statistics for the  deviant groups of stars}\label{tab:6}
  \begin{tabular}{lcccc}
  \hline\hline 
 Group & Quantity   & $D_k$    &  $D_L$ & $D_{Lc}$ \\
  (number)   &                  &   (kpc)    &   (kpc)     &  (kpc)   \\
\hline
Perseus    &  average  &    5.27  & 1.94  & 2.19   \\
  (18)         & standard dev.   &  1.04   &  1.10  &  1.26 \\
				 &  $probability^{\dagger}$ & ---  &  $<10^{-4}$  &  $<10^{-4}$ \\
\hline
    Sgr      &  average  &  3.63   & 2.58  & 3.45  \\
  (21)       & standard dev.   &  0.81   & 0.99  &  2.08 \\
				 & $probability^{\dagger}$  & ---  &  $<10^{-4}$  & 0.66 \\
\hline
reduced   Sgr      &  average  &  3.46   & 2.26  & 2.65  \\
  (17)       & standard dev.   &  0.62   & 0.75  &  1.22 \\
				 & $probability^{\dagger}$  & ---  &  $<10^{-4}$  & 0.02 \\
 \hline
\multicolumn{5}{l}{$^{\dagger}:$ a probability of the Student's t-test for the averages} \\
\multicolumn{5}{l}{ of  two sets of $D_K$ and $D_L$ (or $D_K$  and $D_{Lc}$)} \\
\multicolumn{5}{l}{ being generated by the same distribution function.}   \\
  \end{tabular}
  \end{center}
\end{table} 

The present conclusion strongly depends on the luminosity distance based on the PL relation.
Further examination on the accuracy of the distances will be given in the future paper, which also discuss the validity of the
optical proper motions for these objects. In this paper, we have given a preliminary analysis based on the obtained new radial velocities 
 for a set of optical red variables, and have shown that they provide useful information on the kinematic of the stars in the Galaxy.
 
\section{Summary}
We have observed 379 red variables in the SiO maser lines, obtaining 229 (220 new) detections.
Accurate  radial velocities of the detected sources are used for investigating the kinematics of stars 
in the Solar neighborhood.  Most of the observed stars locate within 3 kpc from the Sun
according to luminosity distances.
The longitude-velocity diagram of the sample shows high number densities
of stars  in two regions. The estimated luminosity distances suggest that these groups of stars 
spatially collocate with the Perseus and Sagittarius spiral arms. 
The result of the VLBI parallax measurements of the objects in these spiral arms 
seems to be consistent with the present data. 
However, at the current moment,  the number of objects observed with VLBI is too small
to conclude the membership of objects to the Galactic spiral arms except for the objects in the Perseus arm. 
In addition, we found a group of stars deviated by more than 40 km s$^{-1}$ in a Galactic longitude range of 20--40$^{\circ}$,
which are likely to be members of the Hercules moving group.  Proper motion data are essential
to reveal the 3d motions of these stars, and the discussion based on the
optically obtained proper motions will be given in a forthcoming paper.

\

We thank Dr. J. Nakashima, Univ. Hong Kong,  for reading the manuscripts and useful comments.
This research was partially supported by a Grant-in-Aid for Scientific Research from
Japan Society for the Promotion of Sciences (20540234).
This research made use of the SIMBAD and VizieR databases operated at CDS, 
Strasbourg, France, and as well as use of data products from 
Two Micron All Sky Survey, which is a joint
project of the University of Massachusetts and Infrared Processing 
and Analysis Center/California Institute of Technology, 
funded by the National Aeronautics and Space Administration and
National Science foundation.

\section*{Appendix. 1. SiO maser spectra and short notes on individual objects}
We show the SiO maser spectra (the $J=1$--0 $v=1$ and 2 transitions) for detections in figures 8a--8m. 
We also show the spectra of other SiO maser transitions in figure 8.
Individually interesting objects are noted as follows.
\begin{itemize}

\item $J02201452+7845362$ (=AG Cep): 
This is a relatively bright IRAS source with $F_{12}=34.1$ Jy.  
H$_2$O maser emission has been detected at $V_{\rm lsr}=0.2$ km $^{-1}$ \citep{lew97},
which is considerably shifted from the SiO radial velocity $V_{\rm lsr}=-28$ km s$^{-1}$
measured in the present paper. OH maser searches have been negative \citep{ngu79,lew95}.
The large velocity difference of about 30 km s$^{-1}$ between H$_2$O and SiO maser lines
suggests that this object is likely a water fountain source \citep{ima07}. The central star, AG Cep, has a 
spectral type of M10, and a pulsation period of 403 d in the NSVS catalog, 
which is slightly different from the period of 445 d in the SVS catalog \citep{sam10}.
The IRAS LRS spectra of this source exhibits a very sharp peak at 9.8 $\mu$m (LRS class 26), and
\citet{lit90} classified the shape of silicate feature as "Sil+", which have high maser detection rate.

\item $J01052742+6558594$ (=V888 Cas): The IRAS LRS spectrum of this object (IRAS 01022+6542)  was 
classified as C type (11 $\mu$m SiC feature;  \cite{vol91}), indicating a carbon star \citep{che03}. 
Searches for the 86.2 GHz SiO and 88.6 GHz HCN  emissions with the IRAM 30m telescope were negative \citep{gro02}.
However, we detected SiO masers in this star at  $V_{\rm lsr}=-63$ km s$^{-1}$. This result
suggests that the LRS  feature is a silicate absorption at 10 
 $\mu$m typically seen in oxygen-rich evolved stars. 
 
 \item $J$03542359+1601019 (=UY Tau) and $J$04212541+2015592 (V1110 Tau):  
 These two stars have similar radial velocities,
 45.5 and 45.2  km s$^{-1}$, respectively,  at the same Galactic longitude $l=175^{\circ}$. 
 The data points for these two stars are overlapped  in the longitude-velocity diagram (Figure 5), lying on
 on the broken curve of the Hercules moving group.  In fact, they are at high Galactic latitudes and 
 separated by 8$^{\circ}$ in Galactic latitude  ($b=-28.1$ and $-20.5^{\circ}$). 
 These two are  likely members of the Hercules moving group.
 
 \item $J04402801+301650$ (=V524 Aur): This is a medium bright IRAS source with $F_{12}$=21.1 Jy and IRAS LRS class of 13 (feature less),
but  this star has been slipped out from the past OH/IR and SiO maser surveys probably because of its "blue" MIR color ($C_{12}=-0.451$). 
The period of light variation of this star, 678 d,  indicates that it is a considerably luminous and massive star among these objects.
SiO masers have been detected for the first time in this paper.  The observed SiO radial velocity ($V_{\rm lsr}=-100.1$ km s$^{-1}$) 
of this star at the Galactic coordinates of $(l,b)=(170.6^{\circ}, -10.7^{\circ})$ indicates that this object is kinematically unusual.

\item $J$07314247+4733226 (=DN Lyn):  This star was once recorded as a dwarf nova and catalogued as
a cataclysmic binary\citep{rit03}.  But this is in fact a faint mira \citep{kaz02}, and the status is corrected
in the later version of the catalog of cataclysmic binaries [see on line version of \citet{rit03}].\footnote{
http://heasarc.gsfc.nasa.gov/W3Browse/all/rittercv.html}
This is a faint IRAS source ($F_{12}=3.7$ Jy), but SiO masers are detected at $V_{\rm lsr}=-37.8$ km s$^{-1}$
in the present paper.

\item $J$20125796$+$3214563 (=V557 Cyg):  This star has an extreme radial velocity of $V_{\rm lsr}=53.4$ km s$^{-1}$
at $l=70.4^{\circ}$ in SiO masers. OH and H$_2$O masers have been detected for this star \citep{lew95}.
The longitude-velocity diagram (Figure 6) shows that several other stars also have similar (but slightly lower) radial velocities:
$J$20074663+3117241 ($V_{\rm lsr}=43.9$ km s$^{-1}$) and  $J$20021291+3057556 ($V_{\rm lsr}=31.8$ km s$^{-1}$).
These three stars fall in a circle of 3 degree diameter, and the estimated distances are between 1.5 and 2.3 kpc
 (though the other two stars also fall near there in Figure 6, their distances are much larger).
Because of their Galactic longitudes ($l\sim 70^{\circ}$),   it is likely that these stars are not associated 
with the Sagittarius spiral arm. They move faster than the Galactic rotation by about 50 km s$^{-1}$.  

\item $J$20195560+8816277 (=X UMi):
This star  is interesting for its location very near to the celestial north pole.
We detected SiO masers at $V_{\rm lsr}=-88.9$ km s$^{-1}$; it is 
 unusual as a star at the Galactic coordinates (121.1$^{\circ}$, 26.5$^{\circ}$);
see Figure 5. This star is located near the edge of the Polaris flare cloud in the local spur \citep{hei90}.
The distance to this flaring cloud is not very far from the Sun, possibly  less than 0.5 kpc  ($V_{\rm lsr}\sim -3$ km s$^{-1}$).
But this star is far away from the CO cloud
because the distance is estimated to be 2.6 kpc. 
The pulsation period of this star is 338 d, and the 2MASS $K_s$ magnitude is 4.3. 
IRAS 12 micron flux is 5.3 Jy with  a color index  $C_{12}=-0.37$.
The spectral class is M8--9 \citep{gig96}.  
Therefore, It is likely that  this is a deviant star in the Perseus arm,
but not a star in the Local spur.
\end{itemize}

\section*{Appendix. 2. Distance estimation using the Period--Luminosity relation}
We estimated the luminosity distance from the observed $K$ magnitudes  
using the Period-Luminosity (PL) relation \citep{whi08}, 
\begin{equation}
M_K=-3.51\times [log (P) -2.38] -7.25.
\end{equation}
The uncertainty of this formula is approximately 0.15 mag  for the miras with a period between 150 and 400 d \citep{whi08}.
The correction for stars with $P>400$ d will be discussed in Appendix 3.
The observed $K$ magnitude can be corrected for the interstellar and circumstellar reddening (see equation (1) of \cite{fuj06})
\begin{equation}
K_c= K - A_K/E(H-K)\times [(H-K)-(H-K)_0],
\end{equation}
where we use $A_K/E(H-K)=1.44$ \citep{nis06},
 and $(H-K)_0 $ is given by the empirical relation 
\begin{equation}
(H-K)_0 = 0.420\times log(P) -0.597
\end{equation}
\citep{cat92}.
Then, we can compute the distance from the difference 
between corrected and absolute $K$ magnitudes, $K_c$ and $M_K$, i.e.,
\begin{equation}
(D_L/{\rm pc})=10^{0.2(K_c-M_K)+1}
\end{equation}
Figure 4 is a histogram of the derived distances using PL relation  for the SiO detections and no detections.
About 80\% of the objects are at the luminosity distances below 3 kpc.

Accuracy of the obtained luminosity distance depends on 
two factors:  reliability of the measured pulsation period and errors in the average $K$ magnitude for a variable star. 
To check the reliability of the periods given by the NSVS catalog, we have cross-correlated the NSVS
periods with those of the SVS catalog. Though the two periods derived from the NSVS and SVS catalogs 
coincide well for the medium-period objects ($P<600$ d) in general,
the coincidence becomes worse for the longer period stars.
Therefore, we have to be careful to derive the distance based on the period given by the NSVS catalog.
The present sample involves not only miras but semi-regulars too. 
Though 75 percent of stars in the present sample are of the variability type of mira,
20 percent of stars are of semi-regular type and 5 percent of stars are of other type
(the latter two types are noted by symbols "$\dagger$" and  "$\ddag$" in the second column of table 4). 
Moreover, some miras with SiO masers occasionally exhibits a pulsation in a first overtone mode
\citep{ita06}. Although  it has been argued that semi-regulars may follow to  a $P$--$M_K$ relation  different from miras
(for example, \cite{bed98}),  the current understanding attributes this phenomenon to the
multiplicity of pulsation modes \citep{tab10}.  For a certain percentage of stars with $P<250$ d follows
to the $P$--$M_K$ relation with almost the same slope but approximately one magnitude brighter than 
the standard sequence of the $P$--$M_K$ relation (sequence C in  a $P$--$M_K$ diagram;
\cite{ita04,tab10}).  However, in the present analysis, we have used  the single $P$--$M_K$ relation (1) 
for all of the observed stars, and estimated the distances. This is because it is hard to specify
the pulsation mode for a particular star with $P<250$ d whether or not it is of higher overtone modes.
Fortunately, the SiO detection rate rapidly decreases for stars with a period below 250 d. 
Furthermore, semi-regulars have a low SiO detection rate ($\sim 13$ \% in the present semi-regular  sample; 
see also \cite{alc90}). In the present sample,  only 18 stars with $P<250$ d were detected in  all the 229 SiO detections.
Therefore, the error in the distance in the SiO detection sample
is not severe. For the 18 Perseus group of stars, no semiregular was involved. 
For the 21 Sgr group of stars, one semiregular with $P=351$ d, which is likely in the sequence C,  
was involved. Therefore, the multiplicity of the $P$--$M_K$ relation for the short period variables
does not affect the discussions made in section 3.

The K-band amplitude of pulsation reaches to 0.8 magnitude for miras \citep{whi08}.
The 2MASS $K_s$ magnitude, which was measured at a single epoch, may differ
from the average value in a pulsation period by about 1 mag [e.g., figure 11 of \cite{mes04}]. 
Furthermore for bright stars with $K<4$, the 2MASS magnitude involves relatively large uncertainty
 (up to 0.4 mag) \citep{cut03}.
Therefore, we deduce that the derived luminosity distance may involve uncertainty of a factor of about 2
for individual objects. However, we expect that the uncertainty do not produce
severe systematic shift in the distance scale and the average value for a certain number of stars is meaningful. 
Therefore, we believe that the uncertainty  of the distance do not mislead the discussion made in the present paper.
\begin{figure}
  \begin{center}
    \FigureFile(80mm,50mm){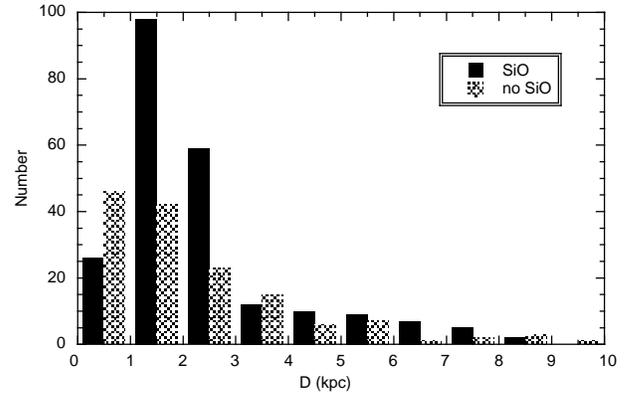}
  \end{center}
  \caption{Histogram of corrected luminosity distances 
  computed using the PL relation (equation 5) for $P>400$ d . 
  The filled and shaded area indicates the SiO detection and non detection.
The average distance is  $2.4\ (\pm 1.8)$ kpc for SiO detections and $2.8\ (\pm 4.2)$ kpc for no detections,
 where the parenthetic number is a standard deviation.   
}\label{fig: hist-distHBB}
\end{figure}

\section*{Appendix. 3. Distance correction for the stars with $P> 400$ d.}
It has been argued that the long-period variables with a period longer than 400 d  lie systematically above
a linear extrapolation of the PL relation of the miras with $100<P<400$ d [see a nice summary
on this problem given by \citet{fea09}].  \citet{whi03} concluded that all the luminous stars found by
the early investigation of \citet{hug90} in the Large Magellanic Cloud (LMC) follow 
an extrapolation of the PL relation except a few stars under the Hot Bottom Burning (HBB) stage.
Because Lithium is overabundant in many OH/IR (and SiO maser) stars \citep{gar07}, it is very likely that
the present sample of SiO maser sources is contaminated by the luminous HBB stars. 
Therefore, we re-estimated the distances of stars introducing the PL relation with a steeper gradient at $P>400$ d.  
 We use for stars with $P>400$ d,
 \begin{equation}
M_K=-6.850\times [log (P) -2.6] -8.406,
\end{equation}
and
\begin{equation}
(H-K)_0 = 1.271 \times [ log(P) -2.6] + 0.271.
\end{equation}
These equations are derived based on the $JHK$ band observations of  the long period variables  in the range $2.6< log(P)<2.95$  
in the LMC \citep{ita11}. The apparent $H$ and $K$ magnitudes in LMC are converted to $M_K$  with a distance modulus of 18.5 for LMC. 
Because the circumstellar extinction is negligibly small for these LMC O-rich stars at $log(P)<2.95$ in their study,
the linear fits of $H$ and $K$ against log(P) in their paper represent the H and K magnitudes without extinction.
Therefore, the difference between the $H$ and $K$ linear fits directly gives $(H-K)_0$. 
The uncertainty of the linear fit in K is deduced to be about 0.36 mag (if we assume the deviation in K is the same 
as that in the LMC; \cite{ita11}).
From the above equations, we can compute the final correction factor for the distance for the stars with $P>400$ d as
\begin{equation}
D_{Lc}/D_L=10^{0.913\times  [ log(P) -2.6]}.
\end{equation}
where $D_{Lc}$ and   $D_{L}$ are the corrected luminosity distance for $P>400$ d 
and the luminosity distance given in Appendix 2 (equation 1), respectively.
The correction factor increases up to about a factor of 2 for $P=850$ d.
We gave the corrected distance for the Perseus and Sgr groups of stars
in parenthetic number at the 9th column of table 5. 
The histogram of the luminosity distances 
in the present sample, which is corrected for all of the $P>400$ d stars,  is shown in Figure 7. 
 
The PL relation for the $P<400$ d stars  does not exhibit a significant offset 
between the LMC and our Galaxy \citep{whi08}. Therefore it is reasonably expected 
that,  for the longer period stars  ($P>400$ d),  the same relation holds both in the LMC and  in our Galaxy,  
except for a special environment such as the Galactic center (e.g.,  \cite{ort02}).

\vspace{3cm}
{\Large \bf A full version of this paper including  Figure  8a--8m, and Figure 9 (SiO maser spectra for detections) are available at \\
http://www.nro.nao.ac.jp/$\sim$lib\_pub/report/data/no680.pdf}.

\clearpage
\setcounter{table}{0}
\clearpage
\tabcolsep 0.5pt
\begin{longtable}{crrrrrrrrr}
  \caption{Observational results of SiO Masers.} 
\hline  \hline
 &  \multicolumn{4}{c}{\underline{  SiO $J=1$--0 $v=1$ maser line  }} & \multicolumn{4}{c}{\underline{  SiO $J=1$--0 $v=2$ maser line  }} & \\
 2MASS name &  $T_a$  &   $V_{\rm lsr}$  & L.F.  &  rms  &      $T_a$  &    $V_{\rm lsr}$ & L.F.  &  rms & obs. date \\  
 {\tiny  (J---)}  &  {\tiny (K)} & {\tiny (km s$^{-1}$)} & {\tiny (K km s$^{-1}$)} &  {\tiny (K)} & {\tiny (K)} & {\tiny (km s$^{-1}$)} & {\tiny (K km s$^{-1}$)} &  {\tiny (K)} & {\tiny (yymmdd.d)}\\
\hline
\endfirsthead
\hline \hline
 &  \multicolumn{4}{c}{\underline{  SiO $J=1$--0 $v=1$ maser line  }} & \multicolumn{4}{c}{\underline{  SiO $J=1$--0 $v=2$ maser line  }} & \\
 2MASS name &  $T_a$  &   $V_{\rm lsr}$  & L.F.  &  rms  &      $T_a$  &    $V_{\rm lsr}$ & L.F.  &  rms & obs. date \\  
  {\tiny  (J---)}  &  {\tiny (K)} & {\tiny (km s$^{-1}$)} & {\tiny (K km s$^{-1}$)} &  {\tiny (K)} & {\tiny (K)} & {\tiny (km s$^{-1}$)} & {\tiny (K km s$^{-1}$)} &  {\tiny (K)} & {\tiny (yymmdd.d)}\\
\hline
\endhead
\hline
\endfoot
\hline
\multicolumn{10}{l}{ "$^{\dagger}$" indicates additional detections in the $^{28}$SiO  $v=0$ or 3, or $^{29}$SiO $v=0$ $J=1$--0  line. see table 3.}\\
\multicolumn{10}{l}{$^1$ previous detection of the SiO $J=2$--1 $v=1$ line \citep{hai94}. }\\
\multicolumn{10}{l}{$^2$ previous detection of the SiO $J=1$--0 $v=1$ and 2 line by \citet{deg10},  $^3$ \citet{deg04}, }\\
\multicolumn{10}{l}{$^4$ \citet{jia99}, $^5$ \citet{cho96}$^6$,  \citet{zuc79}, $^7$\citet{hal90}, $^8$ \citet{kim10} }\\
\endlastfoot
%
00100914$+$5452343  & 0.403 & $-$34.6 & 0.950 & 0.062 & 0.415 & $-$35.9 & 1.695 & 0.068 & 100225 \\
00163648$+$6601104  & 0.283 & $-$60.9 & 0.539 & 0.058 & 0.249 & $-$60.5 & 0.444 & 0.056 & 090309 \\
00202547$+$6947567  & 0.617 & $-$49.5 & 2.026 & 0.082 & 0.938 & $-$49.6 & 2.752 & 0.094 & 100225 \\
00251001$+$7008516  & 1.195 & $-$43.3 & 5.694 & 0.057 & 1.365 & $-$41.3 & 4.394 & 0.067 & 100225 \\
00365942$+$6308016  & 2.428 & $-$58.5 & 11.692 & 0.143 & 2.297 & $-$56.0 & 12.320 & 0.139 & 090308 \\
00382283$+$8021250  & 0.340 & 2.9 & 0.639 & 0.074 & --- & --- & --- & 0.076 & 100321 \\
00452805$+$7550219  & 0.607 & $-$83.1 & 1.998 & 0.058 & 0.438 & $-$82.6 & 1.883 & 0.065 & 100225 \\
00480999$+$5334010  & --- & --- & --- & 0.051 & 0.244 & $-$42.0 & 0.307 & 0.063 & 100225 \\
\end{longtable}
\clearpage
\tabcolsep 2pt
\begin{longtable}{cccc}
  \caption{Negative results for the SiO $J=1$--0 $v=1$ and 2 lines.} 
\hline  \hline
              & \underline{$v=1$}  &  \underline{$v=2$} &  \\
2MASS name   &  rms  &   rms & obs. date \\  
(J---)  & {\tiny (K)} &  {\tiny (K)} & {\tiny (yymmdd.d)}\\
\hline
\endfirsthead
\hline \hline
              &  \underline{$v=1$}  &  \underline{$v=2$} &  \\
2MASS name  &  rms  &   rms & obs. date \\  
  (J---) & {\tiny (K)} &  {\tiny (K)} & {\tiny (yymmdd.d)}\\
\hline
\endhead
\hline
\endfoot
\hline
\endlastfoot
00074306$+$7414113  & 0.070 & 0.077 & 100225 \\
00360130$+$5754427  & 0.092 & 0.087 & 090309 \\
00372236$+$5342117  & 0.081 & 0.098 & 100225 \\
00500597$+$6445351  & 0.077 & 0.090 & 100225 \\
01003486$+$5536221 & 0.092 & 0.097 & 101113 \\
01103043$+$4506115  & 0.092 & 0.097 & 100412 \\
02195542$+$5328072  & 0.086 & 0.089 & 100412 \\
02302753$+$6231456  & 0.108 & 0.102 & 090309 \\
\end{longtable}
\tabcolsep 2pt
\begin{longtable}{ccccccc}
  \caption{Observational results for additional SiO lines} 
\hline  \hline
2MASS name &  transition & $Ta$  &   $V_{\rm lsr}$  & L.F.  &  rms   & obs. date \\  
  (J---) &   & {\tiny (K)} & {\tiny (km s$^{-1}$)} & {\tiny (K km s$^{-1}$)} &  {\tiny (K)}  & {\tiny (yymmdd.d)}\\
\hline
\endfirsthead
\hline \hline
2MASS name &  transition & $Ta$  &   $V_{\rm lsr}$  & L.F.  &  rms   & obs. date \\  
 (J---) &   & {\tiny (K)} & {\tiny (km s$^{-1}$)} & {\tiny (K km s$^{-1}$)} &  {\tiny (K)}  & {\tiny (yymmdd.d)}\\
\hline
\endhead
\hline
\endfoot
\hline
\endlastfoot
00504329$+$4630307 & $^{29}$SiO $v=0$ $J=1$--0 &  0.382  & $-$34.5 &  0.739 & 0.086  & 100222 \\
01070453$+$4924467 & $^{28}$SiO $v=3$ $J=1$--0 & 0.316  & $-$57.7  &  0.445 & 0.081  & 100222  \\
05572394$+$4822417 & $^{29}$SiO $v=0$ $J=1$--0 & 0.521  &   2.2   &   0.835 &  0.084 & 100222  \\
05572394$+$4822417 & $^{28}$SiO $v=3$ $J=1$--0 & 1.629 &    2.8   &   4.406 &   0.105  & 100222  \\
05581447$+$5002407 & $^{29}$SiO $v=0$ $J=1$--0 &  0.305 & $-$43.2 &  1.012  & 0.070  &  100222  \\
14045992$-$3529505  & $^{28}$SiO $v=0$ $J=1$--0 &  0.438 & $-$1.3   & 2.274   & 0.108  &  100223 \\
18272417$-$0100456  & $^{29}$SiO $v=0$ $J=1$--0 &  0.728 &  $-$7.3  & 2.443  & 0.153   &  100315  \\
18541840$-$0648564  & $^{29}$SiO $v=0$ $J=1$--0 &  0.595 & $-$20.8  & 1.066 & 0.116  &   100511  \\
19261909$+$1640323 &  $^{29}$SiO $v=0$ $J=1$--0 &  0.374 &  9.6    &  0.723  & 0.087  &  090308 \\
22071622$+$1153158 & $^{29}$SiO $v=0$ $J=1$--0  &  0.836 &  23.3  &  0.542  & 0.095  & 090309  \\
22071622$+$1153158 & $^{28}$SiO $v=3$ $J=1$--0 &  4.141 &  23.2  &  6.127  &  0.125   & 090309\\
22071622$+$1153158 & $^{28}$SiO $v=0$ $J=1$--0 &   0.334 & 19.1 &   0.505 &  0.104  & 090309 \\
\end{longtable}
\tabcolsep 1pt
\begin{longtable}{lrrrrrrcl}
  \caption{Infrared properties of the observed objects} 
\hline  \hline
2MASS name &  Period & $K$  &  $H-K$ &  $F_{12}$ & $C_{12}$  & $V_{\rm lsr}(SiO)$ & IRAS/MSX & comment \\
     (J---)     &      (d)   & mag. &             &        (Jy)       &                    &  {\tiny (km s$^{-1}$)} &  name  & \\  
\hline
\endfirsthead
\hline \hline
2MASS name &  Period & $K$  &  $H-K$ &  $F_{12}$ & $C_{12}$  & $V_{\rm lsr}(SiO)$ & IRAS/MSX  & comment\\
     (J---)   &      (d)   & mag. &             &        (Jy)       &                    &  {\tiny (km s$^{-1}$)} &name  &  \\  
\hline
\hline
\endhead
\hline
\endfoot
\hline
\multicolumn{9}{l}{$^{\dagger}$: semiregular variables (SR+L), $^{\ddag}$: the other type (MISC);  $^{\sharp 1}$ G038.7780$-$01.1228, $^{\sharp 2}$ G037.3380$-$02.4122,} \\
\multicolumn{9}{l}{$^{\sharp 3}$ G082.5317$+$01.4850,  $^{\sharp 4}$  G084.2104$-$01.8846, $^{\sharp 5}$ G071.9088$-$34.2583, $^{\sharp 6}$  G084.2104$-$01.8846, } \\
\multicolumn{9}{l}{  $^{\sharp 7}$ G262.5919+18.5173, $^{\sharp 8}$ G239.6370+55.6395, $^{\sharp 9}$ G182.7804+72.0228,  $^{\sharp 10}$ G055.3416-64.3560} \\
 \multicolumn{9}{l}{$^{\flat}$ In IRAS point source reject catalog \citep{bei89}. \hspace{5mm} Reference: $^{1}$ \citet{lew97},} \\
	
\multicolumn{9}{l}{ $^{2}$ \citet{cro83}, $^{3}$ \citet{eng88}, $^{4}$ \citet{ede88}, $^{5}$\citet{eng96} ,} \\
\multicolumn{9}{l}{$^{6}$ \citet{lew94}, $^{7}$ \citet{lew90}, $^{8}$ \citet{lew88}, $^{9}$ \citet{ben96}, } \\
\multicolumn{9}{l}{ $^{10}$ \citet{hal90}, $^{11}$ \citet{par03},$^{12}$ \citet{cho10}, $^{13}$ \citet{lew95}} \\
\endlastfoot
00074306$+$7414113 & 476 & 4.64 & 1.12 & 7.4 & $-$0.492 &  & 00050$+$7357 & \\
00100914$+$5452343 & 730 & 3.04 & 0.70 & 15.2 & $-$0.377 & $-$35.2 & 00075$+$5435 & \\
00163648$+$6601104 & 330 & 4.24 & 0.67 & 2.5 & $-$0.286 & $-$60.7 & 00138$+$6544 & \\
00202547$+$6947567 & 425 & 2.86 & 0.86 & 18.6 & $-$0.406 & $-$49.5 & 00176$+$6931 & \\
00251001$+$7008516 & 454 & 2.33 & 0.73 & 25.0 & $-$0.330 & $-$42.3 & 00222$+$6952 & \\
00360130$+$5754427 & 301 & 4.81 & 0.44 & 2.1 & $-$0.450 &  & 00331$+$5738 & \\
00365942$+$6308016 & 730 & 3.07 & 0.80 & 50.4 & $-$0.415 & $-$57.2 & 00340$+$6251 & H$_2$O$^{2,3}$\\
00372236$+$5342117 & 176$^{\dagger}$ & 2.86 & 0.61 & 8.2 & $-$0.439 &  & 00345$+$5325 & \\
\end{longtable}
%

 \clearpage
 \setcounter{table}{4}
\begin{longtable}{cccccccccc}
  \caption{Optical and infrared properties of Perseus and Sgr groups of stars.} 
\hline  \hline
2MASS name   &  Period  &  type & Rmag & Amp(R) & Kmag & e(K) & flg(K) & $D_L$ ($D_{Lc}$)$^{\sharp}$ & $D_K$ \\  
(J---)  & (d) &   &  (mag) & (mag) & (mag) & (mag) & & (kpc) & (kpc) \\
\hline
\endfirsthead
\hline \hline
2MASS name   &  Period  &  type & Rmag & Amp(R) & Kmag & e(K) & flg(K) & $D_L$ ($D_{Lc}$)$^{\sharp}$ & $D_K$ \\  
(J---)  & (d) &   &  (mag) & (mag) & (mag) & (mag) & & (kpc) & (kpc) \\
\hline
\endhead
\hline
\endfoot
\hline
\multicolumn{10}{l}{$^{\flat}$: reliable estimate of the photometric error could not be determined ($>8.0$).}\\
\multicolumn{10}{l}{$^{\sharp}$: parenthetic number is the distance corrected for the HBB stars with $P>400$ d.}\\

\endlastfoot
\multicolumn{10}{c}{Perseus group} \\
00163648$+$6601104 & 330 & M  & 11.816 & 2.786 & 4.241 & 0.059 & E & 2.2 &  5.5 \\
00202547$+$6947567 & 425 & M  & 10.993 & 2.651 & 2.858 & 0.246 & D & 1.2 (1.3)&  4.3 \\
00251001$+$7008516 & 454 & M  & 11.196 & 2.933 & 2.325 & 0.260 & D & 1.1 (1.3)&  3.6 \\
00365942$+$6308016 & 730 & M  & 10.854 & 2.606 & 3.073 & 0.280 & D & 2.2 (3.9)&  5.1 \\
00480999$+$5334010 & 298 & M  &  9.469 & 1.198 & 2.556 & 0.362 & D & 0.9 &  3.5 \\
01052742$+$6558594 & 341 & M  & 10.666 & 2.690 & 2.722 & 0.350 & D & 1.1 &  5.8 \\
01070453$+$4924467 & 405 & M  & 10.426 & 2.414 & 3.184 & 0.350 & D & 1.6 (1.6)&  5.1 \\
\end{longtable}

\end{document}